\documentstyle[prl,aps]{revtex}
\font\trm=cmr10
\font\tbf=cmbx7
\font\latin=cmsy10
\def\npi{\hbox{\latin N$_\pi$}}
\def\ns{\hbox{\latin N$_S$}}

\begin{document}
\draft
\twocolumn[\hsize\textwidth\columnwidth\hsize\csname@twocolumnfalse%
\endcsname


\title{{\hfill\trm SU-ITP 97/15, cond-mat/9705049\bigskip\\}
Is the $\pi$-particle responsible for the 41 meV peak in 
YBa$_2$Cu$_3$O$_7$?}

\author{Martin Greiter}
\address{Department of Physics, Stanford University, Stanford, CA 94305,
greiter@quantum.stanford.edu}

\date{
May 6, 1997}

\maketitle

\begin{abstract}
It is argued that there is no low-energy resonance associated
with the $\pi$ operators introduced by Demler and Zhang.  This implies
that the Hubbard model does not possess an approximate $SO(5)$ 
symmetry generated by these operators.  Recent finite-size studies
are re-interpreted accordingly. 
\end{abstract}

]

1.\
Two years ago, Demler and Zhang\cite{eugene} proposed a new
collective mode for the positive $U$ Hubbard model, the so-called
$\pi$ resonance.  They calculated the energy of this mode within 
the T-matrix approximation, and found 
\begin{equation}
\omega_0 \approx {J\over 2} (1-n)-2\mu ,
\label{eq:oo}
\end{equation}
where $J\approx t^2/4U$ is the antiferromagnetic coupling between
nearest neighbors,
$n<1$ the electron density and $\mu$ the ``chemical potential
measured from half filling''\cite{meix} (by which, I presume, they mean the
difference between the chemical potential at electron density $n$ and
at half filling ($n=1$), as 
there is no meaning in measuring a chemical potential from a particular
reference point).  They then interpreted the sharp magnetic resonance
peak at 41 meV 
observed by inelastic neutron scattering\cite{na,nb,nc,nd}
in the superconducting phase of YBa$_2$Cu$_3$O$_7$ 
in terms of this resonance, and identified $\omega_0$ 
at the appropriate electron density with the observed resonance
frequency. 

2.\
In order to explain why I do not agree with the reasoning leading to
this identification, let me first review the essential steps in their
derivation\cite{eugene}.  The $\pi$ operator is given by
\begin{equation}
\pi_d^\dagger =\sum_{\hbox{\tbf k}} (\cos k_x -\cos k_y) 
c^\dagger_{\hbox{\tbf k}+\hbox{\tbf Q}\uparrow } 
c^\dagger_{-\hbox{\tbf k}\uparrow }
\ \ \ \ {\bf Q}\equiv (\pi,\pi).
\label{eq:pi}
\end{equation}
and is argued to be an approximate eigenoperator of the 
$t$-$J$-$U$-$\mu$ Hamiltonian
\begin{eqnarray}
H &=& - t\sum_{\langle ij\rangle\, \sigma} c^\dagger_{i\sigma}c_{j\sigma}
+ {J\over 2}\sum_{\langle ij\rangle} {\bf S}_i\cdot{\bf S}_j +\nonumber \\
&+& U \sum_i n_{i\uparrow}n_{i\downarrow} 
- \mu \sum_{i\sigma} c^\dagger_{i\sigma}c_{i\sigma},
\label{eq:tju}
\end{eqnarray}
where electron density is determined by the chemical potential.  
This property they deduce from the commutator
\begin{equation}
\lbrack H,\pi_d^\dagger\,\rbrack \approx 2(Un_\downarrow -\mu)\pi_d^\dagger
\,+\,\hbox{term linear in $J$},
\label{eq:com}
\end{equation}
where $n_\downarrow ={1\over N}\sum n_{i\downarrow }$ is the average density
of down-spin electrons.  Careful scrutiny\cite{eugene} 
then reveals that the
term linear in $J$ can be approximated by ${J\over 2}(1-n) \pi_d^\dagger$.
Demler and Zhang obtained (\ref{eq:oo}) from (\ref{eq:com}) by
asserting that ``the leading contribution to the chemical potential
is [in the large $U$ limit] given by $Un/2$'', and that the first term
in (\ref{eq:com}) ``cancels in the leading order in $U$''\cite{eugene}.
With this I disagree.

First of all, let us go back and find the source of the mistake.  In 
an earlier paper\cite{zhang}, Zhang correctly observes that the 
particle hole symmetry of the Hubbard model implies
\begin{equation}
\mu(n)=U-\mu(2-n) 
\end{equation}
and incorrectly infers that $\mu =U/2$ at half filling\cite{reve}%
---incorrectly because the chemical potential of the positive-$U$ 
Hubbard model is (in general) discontinuous
at \hbox{$n=1$,} with a discontinuity which becomes more and more pronounced
as $U$ tends to infinity: 
\begin{equation}
\mu(1+\varepsilon)-\mu(1-\varepsilon) = U-2\mu(1-\varepsilon),
\end{equation}
where $\varepsilon$ is a positive infinitesimal and $\mu(1-\varepsilon)$
approaches a constant as $U\to\infty$. 

The origin of this discontinuity is best understood in the same
limit:  up to half filling, the ground state is to zeroth order
composed of configurations without doubly occupied sites;
doubly occupied sites only enter through virtual processes.
The chemical potential in this regime will therefore approach a
(density dependent) constant in the large $U$ limit.
Once we exceed half filling, however, the ground state must
necessarily contain as many doubly occupied sites as there are electrons
exceeding half filling, with a zeroth order energy cost of $U$
associated with each.
The leading contribution to the chemical potential in that regime
is thus given by $U$.

The chemical potential for the large $U$ Hubbard model with $n<1$ is
consequently much smaller than $Un/2$, and the energy of the $\pi$ particle
\begin{equation}
\omega_0 \approx 2 U n_\downarrow - 2\mu (n) + {J\over 2}(1-n),
\label{eq:coc}
\end{equation}
is of order 2 eV.  This particle cannot
possibly account for the magnetic resonance peak observed at 41 meV.

3.\
The physical reason the energy of the $\pi$ excitation is of order
$U$ is trivial.  The $\pi$ operator places two up-spin
electrons with a center of mass momentum of $(\pi,\pi)$ on two
neighboring lattice sites, with an amplitude insensitive to whether
these sites are occupied by holes (that is unoccupied) or by
down-spin electrons.  (If one of the sites is occupied by
an up-spin electron, the amplitude vanishes due to the Pauli
principle.)  Therefore, as the $\pi$ operator acts on the ground
state for the Hubbard model with electron density $n<1$, there
is a large amplitude to find doubly occupied sites.  Those give rise
to the first term in (\ref{eq:coc}).  

Unlike the $\eta$ excitation proposed by Yang\cite{yang} and
further elaborated by Zhang\cite{scz}, the $\pi$ excitation is
hardly a resonance.  The $\eta$ operator places a pair electrons
with opposite spin on the same lattice site, and as the Pauli
principle excludes the remaining electrons from this site, there is no
scattering between this pair and any of the other electrons by
the Hubbard interaction $U$.  The situation is different for
the $\pi$ operator, which places a pair of up-spin electrons
on neighboring sites, and thereby allows $U$ to scatter any
down-spin electron in the liquid off this pair.
This scattering induces an almost instantaneous decay of the
$\pi$ excitation, with a decay rate of order $U$.

\smallskip
4.\
The fact that there is no low-energy resonance associated with
the $\pi$ operators implies that the $\pi$ operators do not generate
an approximate $SO(5)$ symmetry for the Hubbard model\cite{sofive}.  
\smallskip

5.\
In light of these considerations, it comes as a surprise when
Meixner {\it et al.}\cite{meix} present finite size studies which
``verify that the $\pi$ operators are approximate
eigenoperators of the Hubbard model'' and ``lend support to
an interpretation of the recent neutron scattering peaks'' in
terms of the $\pi$ excitation.
In figures 1(b) and 1(f), 2(b) and 2(f) of their manuscript, they
compare the $\pi$-$\pi$ correlation function
\begin{equation}
\pi_d^+(\omega ) = -{1\over \pi}\Im \langle\Psi_0^{N-2}\vert \pi_d
{1\over \omega-H+E_0^N+i\varepsilon} \pi_d^\dagger \vert\Psi_0^{N-2}\rangle.
\label{eq:pipi}
\end{equation}
with the dynamical spin-spin correlation function
\begin{equation}
\chi_{\hbox{\tbf Q}}^+(\omega ) = 
-{1\over \pi}\Im \langle\Psi_0^N\vert S_{\hbox{\tbf Q}}^-
{1\over \omega-H+E_0^N+i\varepsilon} S_{\hbox{\tbf Q}}^+ \vert\Psi_0^N\rangle.
\label{eq:spsp}
\end{equation}
where $\Im$ denotes the imaginary part and $H$ the $t$-$U$
Hubbard Hamiltonian (which does not contain a chemical potential term).
The $\pi_d$ operator is as defined in (\ref{eq:pi}), and
\begin{equation}
S_{\hbox{\tbf Q}}^+ = \sum_{\hbox{\tbf k}} 
c^\dagger_{\hbox{\tbf k}+\hbox{\tbf Q}\uparrow } c_{\hbox{\tbf k}\downarrow }
\ \ \ \ S_{\hbox{\tbf Q}}^-={S_{\hbox{\tbf Q}}^+}^\dagger
\ \ \ \ {\bf Q}\equiv (\pi,\pi)
\label{eq:sq}
\end{equation}
is the spin density wave operator.
For a 10 site Hubbard cluster with N=10, they find that both these correlation
functions (as well as a mixed correlation function involving both $\pi_d$
and $S_{\hbox{\tbf Q}}^+$) have sharp resonance peaks at the same energy.  
They conclude that this ``clearly demonstrates that the peak in the spin
correlation function is due to a particle-particle intermediate state
$\pi_d^\dagger\vert\Psi_0^{N-2}>$''\cite{meix}. 

6.\
In order to explain why I disagree, let me first review the
reasoning leading to this interpretation\cite{scz}.  For this purpose,
let us assume that the $\pi$ operator was an exact eigenoperator
of $H$, 
\begin{equation}
\lbrack H,\pi_d^\dagger\,\rbrack = \omega_\pi \pi_d^\dagger .
\label{eq:picom}
\end{equation}
Then the only intermediate state which would contribute 
to the $\pi$-$\pi$ correlation function defined in (\ref{eq:pipi}),
\begin{equation}
\pi_d^+(\omega ) = -{1\over \pi}\Im \sum_n 
{\left|\langle\Psi_n^N\vert \pi_d^\dagger \vert\Psi_0^{N-2}\rangle\right|^2
\over \omega -E_n^N+E_0^N+i\varepsilon},
\label{eq:pipia}
\end{equation}
where the sum extends over all the excited states of $H$, would be 
\begin{equation}
\vert\Psi_\pi^N\rangle = 
\npi \pi_d^\dagger \vert\Psi_0^{N-2}\rangle,
\label{eq:psipi}
\end{equation}
where $\npi$ is a normalization constant; we choose $\npi$ real.
Therefore (\ref{eq:pipia}) would reduce to
\begin{equation}
\pi_d^+(\omega ) = {1\over \hbox{\npi}^2}\delta (\omega -\omega_\pi).
\label{eq:pipib}
\end{equation}

In the presence of $d$-wave superconductivity, this resonance would also 
manifest itself in the spin-spin correlation function defined in
(\ref{eq:spsp}),
\begin{equation}
\chi_{\hbox{\tbf Q}}^+(\omega ) = -{1\over \pi}\Im \sum_n 
{\left|\langle\Psi_n^N\vert S_{\hbox{\tbf Q}}^+ \vert\Psi_0^N\rangle\right|^2
\over \omega -E_n^N+E_0^N+i\varepsilon},
\label{eq:spspa}
\end{equation}
as the exact excited state $\vert\Psi_\pi^N\rangle$ 
would yield a singularity at $\omega_\pi$:
\begin{eqnarray}
\chi_Q^+(\omega ) &=& -{1\over \pi}\npi^2 \Im
{\left|\langle\Psi_0^{N-2}\vert \pi_d S_{\hbox{\tbf Q}}^+ 
\vert\Psi_0^N\rangle\right|^2 \over \omega -\omega_\pi +i\varepsilon}+ 
\nonumber \\ &&+\ \hbox{regular at $\omega_\pi$}= \nonumber \\
&=& 4 \npi^2\
\left|\langle\Psi_0^{N-2}\vert \Delta_d
\vert\Psi_0^N\rangle\right|^2 \ \delta(\omega -\omega_\pi) + \nonumber \\
&&+\ \hbox{regular at $\omega_\pi$},
\label{eq:spspb}
\end{eqnarray}
where we have used
\begin{equation}
{1\over 2} \lbrack \pi_d ,S_{\hbox{\tbf Q}}^+ \rbrack
= \sum_{\hbox{\tbf k}} (\cos k_x -\cos k_y) c_{\hbox{\tbf k}\uparrow }
c_{\hbox{\tbf -k}\downarrow }
\equiv \Delta_d .
\label{eq:delta}
\end{equation}
The expectation value
$\langle\Psi_0^{N-2}\vert \Delta_d \vert\Psi_0^N\rangle $
is often used as $d$-wave superconducting order parameter.  

This effect---the manifestation of a particle-particle resonance
in the particle-hole channel in the presence of superconductivity---persists
even if $\vert\Psi_\pi^N\rangle$ is only an approximate
eigenstate.  Hence their interpretation.

7.\
These authors, however, seem to have overlooked that this argument
can be run backwards.  In the presence of superconductivity, a
resonance in the particle-hole channel will manifest itself in
the particle-particle channel.  In particular, if the one-magnon
state
\begin{equation}
\vert\Psi_S^N\rangle = 
\ns S_{\hbox{\tbf Q}}^+ \vert\Psi_0^N\rangle,
\label{eq:psis}
\end{equation}
was an exact eigenstate of $H$,
\begin{equation}
\lbrack H,S_{\hbox{\tbf Q}}^+\rbrack =\omega_S S_{\hbox{\tbf Q}}^+,
\label{eq:scom}
\end{equation}
this resonance would manifest itself in the $\pi$-$\pi$ correlation
function:
\begin{eqnarray}
\pi_d^+(\omega ) &=&4 \ns^2\
\left|\langle\Psi_0^{N-2}\vert \Delta_d
\vert\Psi_0^N\rangle\right|^2 \ \delta(\omega -\omega_S)
+ \nonumber \\ &+& \hbox{regular at $\omega_S$},
\label{eq:pipic}
\end{eqnarray}
Again, the effect persists even if $\vert\Psi_S^N\rangle $ is only
an approximate eigenstate.  This is how I would interpret the numerical 
studies by Meixner {\it et al.}\cite{meix}; 
a magnon manifests itself in the particle-particle channel.

In fact, a simple {\it Gedankenexperiment} can decide between the two
interpretations.
Suppose we destroy the $d$-wave superconductivity in the 10 site
Hubbard cluster by adding a nearest neighbor repulsion of
order $U'=4t^2/U$.  Then the resonances in the two different channels
do no longer have to coincide in frequency; one of them may be shifted,
or might even disappear.  Since the resonances in the spin-spin
correlation function $\chi_Q^+(\omega )$ shown in figures 1(f) and
2(f) of the mentioned manuscript occur at an electron density
$n=1$ (that is one electron per site), a nearest
neighbor repulsion $U'\ll U$ would only induce a higher order correction;
it would not affect the one-magnon resonance significantly.
The nearest neighbor repulsion, however, would be likely to affect the
resonance in the $\pi$-$\pi$ correlation function $\pi_d^+(\omega )$
at $n=0.8$ shown in figures 1(b) and 2(b) of the manuscript.
Hence my interpretation.

\smallskip

I am deeply grateful to Eugene Demler and Shou-Cheng Zhang for 
their generosity in explaining many aspects of their work to me, to 
Stefan Meixner for providing me with an unpublished figure, and
to Bob Laughlin for his critical reading of the manuscript.
This work was supported through NSF grant No.~DMR-95-21888.
Additional support was provided by the NSF MERSEC Program through 
the Center for Materials Research at Stanford University.


\begin{references}
\bibitem{reve}
The reverse of this statement is of course true:
if we choose $\mu=U/2$, the Hubbard cluster will be half filled,
as it will be for any choice of $\mu$
between $\mu(1-\varepsilon)$ and $\mu(1+\varepsilon)$.
The chemical potential plotted as a function of the electron
density is discontinuous at $n=1$, while the electron density
plotted as a function of the chemical potential has a plateau
between $\mu(1-\varepsilon)$ and $\mu(1+\varepsilon)$.
\end{references}
\end{document}